# TSV-integrated Surface Electrode Ion Trap for Scalable Quantum Information Processing

P. Zhao[1,3], J. P. Likforman[2], H.Y. Li[1], J. Tao[3], T. Henner[2], Y. D. Lim[3], W. W. Seit[1], C. S. Tan[3, b)], L. Guidoni[2, a)]

[1]Institute of Microelectronics, Agency for Science, Technology and Research (A*STAR), Singapore 117685
[2]Laboratoire Matériaux et Phénomènes Quantiques, CNRS - Université de Paris, F-75013 Paris, France
[3]School of Electrical and Electronic Engineering, Nanyang Technological University, Singapore 639798

In this study, we report the first Cu-filled through silicon via (TSV) integrated ion trap. TSVs are placed directly underneath electrodes as vertical interconnections between ion trap and a glass interposer, facilitating the arbitrary geometry design with increasing electrodes numbers and evolving complexity. The integration of TSVs reduces the form factor of ion trap by more than 80%, minimizing parasitic capacitance from 32 to 3 pF. A low RF dissipation is achieved in spite of the absence of ground screening layer. The entire fabrication process is on 12-inch wafer and compatible with established CMOS back end process. We demonstrate the basic functionality of the trap by loading and laser-cooling single $^{88}$Sr$^+$ ions. It is found that both heating rate (17 quanta/ms for an axial frequency of 300 kHz) and lifetime (~30 minutes) are comparable with traps of similar dimensions. This work pioneers the development of TSV-integrated ion traps, enriching the toolbox for scalable quantum computing.

Quantum information processing (QIP) platforms recently released from IonQ and Honeywell have highlighted the impressive development of surface electrode ion traps[1,2]. Taking advantage of well-established microfabrication techniques (i.e., MEMS and CMOS), surface electrode ion traps exhibits flexible design, reproducible fabrication and mass production[3]. More importantly, the possible integration with photonics components (i.e., waveguide, grating coupler[4-6] and photodetector[7]) enables precise control and measurement on individual trapped ion. These features make surface electrode ion trap highly promising for large scale QIP. To scale close or even beyond noisy intermediate-scale quantum regime (NISQ, ~100 ions)[8], 1D arrays, 2D arrays[9,10] and quantum charge-coupled device (QCCD) geometry[11,12] have been proposed with evolving electrodes layout complexity. However, this strategy opens significant challenges for electric signal delivery, since conventional bonding wires are not able to access the isolated island-like electrodes, and large number of wires will further impede laser beams.

The previous approach to mitigate this challenge is to employ multilayer metallization, where patterned metal layers and dielectric material are alternatively overlapped beneath surface electrode[13-16]. Small vias through dielectric layers are required to build interconnections between different metal layers, while the metal layer in the bottom is often used as grounding plane to shield the silicon substrate from RF signal. The incorporation of routing leads underneath allows flexible design of surface electrodes geometry. However, this approach also comes with issues. First, as photonics layer is normally laid beneath following metallic layers due to its stringent wafer flatness requirement, the multilayer metallization above will significantly hamper its on-chip integration, and thus degrade the ion trap scalability. In addition, the thick dielectric layer (~10 μm) formation and patterning necessitate non-standard fabrication techniques, limiting the compatibility with large-scale foundry manufacturing[17]. Eventually, the multilayer metallization may increase the coupling parasitic[18], resulting in high RF loss and heating of the device. Recently, the use of via as interconnection of ion trap has also been explored. As reported by N. D. Guise et al[19], polysilicon filled via with diameter of ~50 μm was used for signal delivery. However, due to the poor conductivity of polysilicon, only DC electrodes were contacted with vias and additional platinum silicide contacts were required as ohmic contacts between electrodes and polysilicon. In another work (H. Li et al[20]), a ring trap was contacted with a 100 μm diameter electrical via etched through a 70 μm glass substrate followed by 800 nm gold deposition. However, this customized process curtails the large-scale fabrication compatibility.

In order to overcome these limitations and further boost the scalability, the first Cu-filled TSV integrated ion trap is demonstrated in this work. With intrinsically small resistance, Cu-filled TSV are used as vertical connections between all the electrodes (including RF) and an interposer underneath. The first advantage of this approach, compared to multilayer metallization, is that the small footprint of TSV makes it highly compatible with photonics integration: the small diameter (20 μm) of TSV allows for compact and flexible electrode design. Besides, TSV eliminates the requirement for thick dielectric layer and the patterning of thin SiO2 layer (~3 μm) between electrode and substrate is of less complexity and foundry compatible. Additionally, as bonding wires are assigned to the interposer, the height difference between ion trap and interposer may mitigate laser beam blocking issues.

We designed the TSV integrated ion trap (TSV trap) starting from a previous 5 wire design[21,22] that includes bonding pads (WB, wire bonding trap). As shown in Fig. 1(a), TSV trap geometry only preserves the core region, eliminating the wire bonding pads and the connecting circuits present in the WB trap[23]. The original wire

---

a) luca.guidoni@univ-paris-diderot.fr
b) tancs@ntu.edu.sg

bonding pads are transferred to an interposer, reducing drastically the surface of the electrodes. This minimizes undesired parasitic and mitigate the RF loss issue. The width and length of center three parallel electrodes (RF, GND and RF) are 80 and 2920 μm respectively, while the gap in between is 5 μm. TSV are designed with a diameter of 20 μm and depth of 100 μm. The pitch between TSVs is 100 μm. 10 TSVs are accommodated under RF electrodes, while 6 or 10 TSVs are accommodated under surrounding DC electrodes, depending on the electrode size. Micro bumps with a diameter of ~30 μm are located underneath every single TSV to support ion trap after assembling with interposer. The interposer is designed with a redistribution layer (RDL), acting as a bridge between the ceramic pin grid array package (CPGA) and the TSV trap for electrical signal transmission (Fig. 1(b)). The tested RDL geometry is a sort of extension of ion trap electrodes. Obviously, more complex routing can be implemented where necessary. While the TSV trap is made on a silicon substrate, we choose to use a glass substrate to fabricate the interposer, in order to minimize RF losses and simplify the fabrication process. To evaluate a possible effect of the electrode geometry modification between WB and TSV traps, a finite element modelling (FEM) is conducted. As expected, we find negligible effects on both the trapping height (~76 μm) and trapping depth (~80 meV with a RF amplitude $V_{RF}$ = 200V at 60 MHz).

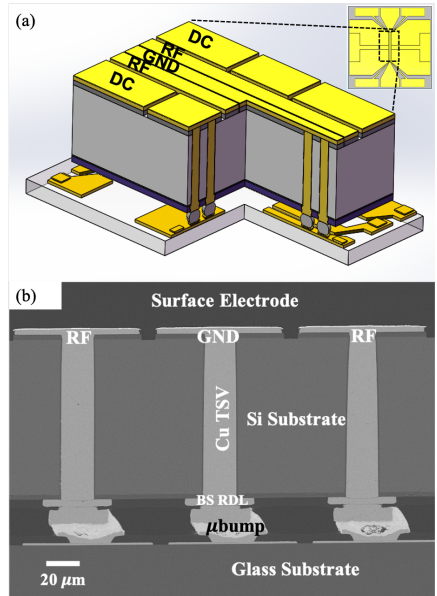

FIG. 1. (a) 3D schematic of TSV-integrated ion trap (TSV trap, not to scale). The geometry is modified from wire bonding (WB) trap (inset). (b) Cross-sectional SEM image of TSV trap. The surface electrode and interposer RDL are interconnected by TSV and micro bumps.

Fabrication processes of both glass wafer (for interposer) and silicon wafer (for ion trap device) are compatible with standard back end process in CMOS foundry, in which 12-inch wafer are used and ~2,000 single dies with different design dimensions are fabricated simultaneously (Fig. 2(a) and (b)). For silicon wafer fabrication, blind vias are first etched through ~2.4 μm $SiO_2$ and 100 μm Si before wafer frontside patterning. To insulate Cu core from silicon, a TSV liner with thickness of 0.75 μm is uniformly deposited onto via sidewall. After Cu filling, two steps of chemical-mechanical polishing are required before and after annealing to completely remove Cu overburden. Next, the patterning of frontside $SiO_2$ is conducted. Note that $SiO_2$ is designed with same geometry as electrodes, in order to avoid undesired dielectric charging issue during ion trapping operation. To form electrodes, lithography defined electroplating of 3 μm Cu with 0.3 μm Au finish is employed. Following that, both wet etching and plasma etching are used to minimize the metal seed residue in the electrode gaps. After temporary bonding with handle wafer, the device wafer is thinned to ~100 μm by grinding and dry etching. Finally, three steps of lithography are conducted in sequence to define RDL, passivation layer and micro bumps on the wafer backside. On the other hand, the fabrication of glass interposer is similar with backside process of ion trap, where three steps of lithography are required, but the last one is to define under bump metallization (UBM) instead of micro bump. After wafer singulation, selected known good dies are bonded together. Eventually, bonded ion trap is assembled into CPGA, as shown in Fig. 2(c).

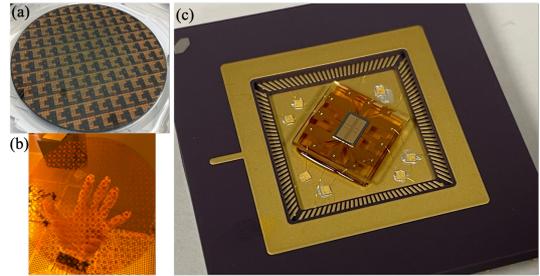

FIG. 2. (a) Fabricated TSV die on 12-inch silicon wafer. (b) Fabricated interposer on 12-inch glass wafer. (c) Packaged TSV trap in a CPGA.

To assess the reliability of TSV liner, a voltage sweep from -200 to 200 V is applied to TSVs. It is found the maximum leakage current for a single TSV is about $7 \times 10^{-12}$ A, suggesting 0.75 μm TSV liner is able to withstand a voltage of 200 V and will not breakdown. Also, a resistance measurement is performed between RF and GND electrodes on both ion trap and glass interposer. The results shows a minimum resistance of ~2 $\times 10^8$ Ω, which is sufficiently large for voltage insulation. In addition, the connection between interposer RDL and corresponding surface electrodes is checked. No open circuit is found, indicating a firm and stable signal transmission path.

One of the major challenges for ion trap with semiconductor substrate is the high RF dissipation[24-26]. To evaluate the RF performance of TSV trap, a WB trap with silicon substrate but no grounding plane is used for comparison. Given that the power loss is proportional to the square of parasitic capacitance of ion trap[21,25], a capacitance-voltage (*C-V*) test is first performed to measure the capacitance between RF and GND electrodes. The result shows that TSV trap has a capacitance of 3 pF, as compared to 32 pF of WB trap. This 90% capacitance decrease can be attributed to the significant reduction of electrode surfaces (from 4.5 to 0.2 mm$^2$, for single RF electrode). A resonance test is then conducted[27]. As shown in Fig. 3(a), TSV trap features a two orders of magnitude RF loss reduction as compared to WB trap. Its resonance peak is close to that of a standard RF capacitor, which suggests that a low RF loss is achieved for TSV trap even though no grounding plane is added.

On-chip S parameter measurement is carried out and the overlapping curves from TSV die (without interposer) and TSV trap (with interposer) in Fig. 3(b) indicate that negligible loss is induced by the incorporation of glass interposer.

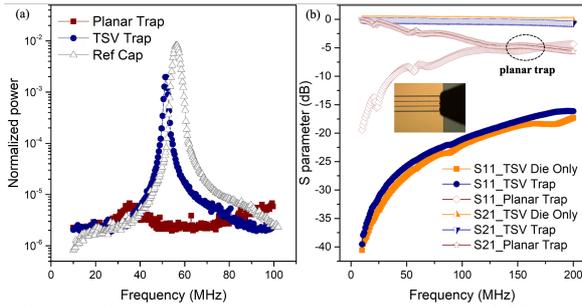

FIG. 3. (a) Resonance test of ion traps packaged into CPGA. The reference curve is from a standard capacitor. (b) On-chip S parameter test of TSV die before and after bonding with glass interposer. The curves from WB trap are also plotted for comparison, showing TSV trap can reduce the insertion loss from -2.2 to -0.1 dB (at 50 MHz). The inset shows a three-pin probe contacting on electrodes surface.

The packaged TSV trap of Fig. 3(c) is tested in a laser-cooled trapped ion setup by loading it with single $^{88}$Sr$^+$ ions. Technical details of the setup are described elsewhere[28,29].

We insert the CPGA in the vacuum cell and then bake at 150°C for one week. The resulting base pressure is approximately $4 \times 10^{-11}$ mbar, demonstrating the excellent ultra-high vacuum compatibility of the packaged ensemble (trap, interposer and micro-bumps). We drive the two RF lines with an impedance-matching toroidal resonant transformer at a frequency $\Omega_{RF}/2\pi$ = 30 MHz. We only test the trap with amplitudes $V_{RF}$ below 120 V. For $V_{RF}$ larger than 85 V we notice a slight increase of the base pressure (of the order of some $10^{-11}$ mbar), probably associated to residual heat dissipation in the trap. We load the trap by two-colour photo-ionizing a thermally sublimated Sr atomic beam[30,31].

The $^{88}$Sr$^+$ ions are Doppler-cooled addressing the $5^2S_{1/2} \rightarrow 5^2P_{1/2}$ transition (711 THz, 422 nm). To avoid optical pumping into the metastable $4^2D_{3/2}$ state we use two additional lasers ("repumpers") addressing the 299~THz $4^2D_{3/2} \rightarrow 5^2P_{3/2}$ transition and the 290 THz $4^2D_{5/2} \rightarrow 5^2P_{3/2}$ transition (incoherent repumping scheme[32]). 711 THz photons scattered by the ion are collected by a home-made objective with numerical aperture of 0.4 and detected by a photomultiplier in photon-counting mode. An overall collection efficiency of $(1.8\pm0.2 \times 10^{-3})$ is measured using sequential acquisitions that require the 275 THz "readout" laser addressing the $4^2D_{3/2} \rightarrow 5^2P_{1/2}$ transition[33].

We also acquire images of the trapped and cooled ions with an electron-multiplier CCD camera (Andor Luca), as shown in Fig. 4(a).

Excess of micromotion in the trap is minimized using a single photon time correlation method[34]. The compensation DC voltages are found to be stable in a week basis. The lifetime of a laser-cooled ion in the trap is of the order of 30 minutes, compatible with the vacuum level.

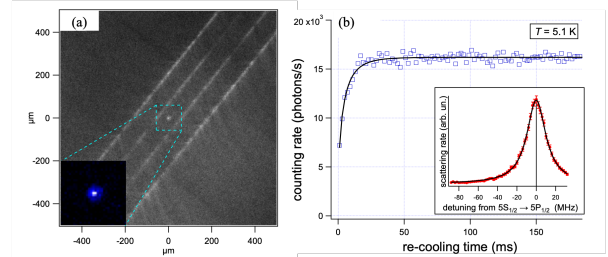

FIG. 4. (a) Image of a single $^{88}$Sr$^+$ laser-cooled ion trapped above the surface of the TSV trap; the image field covers a 1 mm$^2$ area, the trap is illuminated by a light emitting diode. (b) Counting rate as a function of the re-cooling time averaged over 120 realisations for a heating time $t_h$ = 20 seconds (blue squares). The continuous black curve is the best one-parameter fit that allows us to retrieve a temperature increase T=5.1 K. Inset: fluorescence spectrum of the ion as a function of the detuning of a probe beam that scans the cooling transition (red circles); the continuous black curve is a Lorentzian fit that shows that the incoherent repumping scheme leads to a very good two-level atom approximation.

We evaluate the trap performances by measuring the heating rate with the technique of Doppler re-cooling[35].

To this purpose we operate the trap with an axial frequency $\omega_z/2\pi$ = 300 kHz and radial frequencies around $\omega_r/2\pi$ = 2.6 MHz. A sequential acquisition first cools the ion during 500 ms, then switches off the cooling laser for a waiting (heating) time $t_h$, and then switches on again the cooling laser triggering the acquisition of single photons timestamped with arrival times. The two-level-atom approximation needed for the analysis of the acquired data [36] is well fulfilled by the incoherent repumping approach: we show in Fig. 4(b, inset) a single-ion fluorescence spectrum that displays a Lorentzian line-shape. We record single shot and average histograms of the scattered photons as a function of the emission time for different heating times $t_h$. An example of averaged histogram is plotted in Fig. 4(b) with the corresponding fit obtained with the hypothesis of a Maxwell-Boltzmann velocity distribution (T=5.1 K for $t_h$ = 20 s). From the analysis of all the experimental sets the heating rate of the trap is evaluated at 250±15 mK/s that corresponds to 17 axial quanta per millisecond. The calculated figure of merit corresponding to the noise spectral density times the motional frequency[37] is $\omega_z*S_E(\omega_z)$ =1.5 × 10$^{-4}$ (V/m)$^2$. While it does not set a new record, this result favorably compares to non-decontaminated non-cryogenics traps of similar dimensions[38].

In conclusion we demonstrate the successful fabrication in a CMOS foundry and the full functionality of a silicon surface ion trap in which all the electric connections (i.e., DC and RF) are realized with through silicon vias. We load single $^{88}$Sr$^+$ ions into the trap and we find that the heating rate, lifetime and stability are comparable with other traps of similar dimensions. This work pioneers the development of TSV-integrated ion traps, enriching the toolbox for trapped ion based scalable quantum computing. In particular the TSV approach is compatible with photonic circuit integration[4-6], insertion of a ground screening layer to eliminate trap-heating, and in the future could be extended to glass substrates[29].

We would like to thank technical staffs in in Institute of Microelectronics, A*STAR for their technical supports on trap fabrication and packaging. We acknowledge the funding support from A*STAR Quantum Technology for Engineering (A1685b0005). The data that support the findings of this study are available from the corresponding author upon reasonable request.